\documentclass[11pt,superscriptaddress,aps,prd,preprint]{revtex4}

\everymath{\displaystyle}
\usepackage{graphicx}

\usepackage[T1]{fontenc}
\usepackage{amsmath}
\usepackage{amssymb}
\usepackage{graphicx}
\usepackage{xcolor}

\newcommand{\bea}{\begin{eqnarray}}
\newcommand{\eea}{\end{eqnarray}}

\begin{document}

\title{Non-Abelian Gravitoelectromagnetism and applications at finite temperature}

\author{A. F. Santos}
\email{alesandroferreira@fisica.ufmt.br}
\affiliation{Instituto de F\'{\i}sica, Universidade Federal de Mato Grosso,\\
78060-900, Cuiab\'{a}, Mato Grosso, Brazil}

\author{J. Ramos}
\email{jramos@burmanu.ca}
\affiliation{Faculty of Science, Burman University\\
Lacombe, Alberta, Canada T4L 2E5}

\author{Faqir C. Khanna \footnote{Professor Emeritus - Physics Department, Theoretical Physics Institute, University of Alberta\\
Edmonton, Alberta, Canada}}
\email{khannaf@uvic.ca}
\affiliation{Department of Physics and Astronomy, University of
Victoria,BC V8P 5C2, Canada}

\begin{abstract}

Studies about a formal analogy between the gravitational and the electromagnetic fields lead to the notion of Gravitoelectromagnetism (GEM) to describe gravitation. In fact, the GEM equations correspond to the weak field approximation of gravitation field. Here a non-abelian extension of the GEM theory is considered. Using the Thermo Field Dynamics (TFD) formalism to introduce temperature effects some interesting physical phenomena are investigated. The non-abelian GEM Stefan-Boltzmann law and the Casimir effect at zero and finite temperature for this non-abelian field are calculated.

\end{abstract}

\maketitle

\section{Introduction}

The Standard Model (SM) is a non-abelian gauge theory with a symmetry group ${\cal U}(1)\times {\cal SU}(2)\times {\cal SU}(3)$. SM describes theoretically and experimentally three of the four fundamental forces of nature, i.e., the electromagnetic, weak and strong forces. The electromagnetism is a ${\cal U}(1)$ abelian gauge theory which has been tested to a high precision. The generalization of an abelian gauge theory to the non-abelian gauge theory was proposed by Yang and Mills \cite{YM}. The last one describes the electroweak unification and quantum chromodynamics. The electroweak interaction is described by an ${\cal SU}(2)\times {\cal U}(1)$ group and  while the ${\cal SU}(3)$ group satisfies the quantum chromodynamics \cite{Marciano, Marciano1, Brambilla}. 

Gravity is not a part of SM. This implies that the SM is not a fundamental theory that describes all fundamental interactions of nature. In this paper, an extension of non-abelian gravity is discussed. Some applications  of such a theory are developed. The gravitational theory studied here is the Gravitoelectromagnetism (GEM). GEM is an approach based on describing gravity in a way analogous to the electromagnetism \cite{Maxwell,Heaviside1, Heaviside2}. Several studies about GEM theory have been developed \cite{Matte, Campbell1, Campbell2, Campbell3, Campbell4, Braginsky, Jantzen}. These ideas arise from the analogy between equations for the Newton and Coulomb laws and the interest has increased with the discovery of the Lense-Thirring effect, where a rotating mass generates a gravitomagnetic field \cite{12}. Some experiments that study this effect have been developed, such as LAGEOS (Laser Geodynamics Satellites) and LAGEOS 2 \cite{13}, the Gravity Probe B \cite{14} and the mission LARES (Laser Relativity Satellite) \cite{15, 16}.

The GEM theory may be analyzed by three different approaches: (i) using the similarity between the linearized Einstein and Maxwell equations \cite{Mashhon}; (ii) a theory based on an approach using tidal tensors \cite{Filipe} and (iii) the decomposition of the Weyl tensor  ($C_{ijkl}$) into ${\cal B}_{ij}=\frac{1}{2}\epsilon_{ikl}C^{kl}_{0j}$ and ${\cal E}_{ij}=-C_{0i0j}$, the gravitomagnetic and gravitoelectric components, respectively \cite{Maartens}. In this paper the Weyl tensor approach is used. A Lagrangian formulation for GEM is developed \cite{Khanna} and a gauge transformation in GEM is studied \cite{gauge}. Here an extension to non-abelian GEM fields is introduced. Applications of the non-abelian GEM at finite temperature are investigated. The temperature effects are introduced using Thermo Field Dynamics (TFD) formalism.

There are two ways to introduce the temperature effect: (i) using the imaginary time formalism \cite{Matsubara} and (ii) using the real time formalism \cite{Sch, Keld, Umezawa1, Umezawa2, Umezawa22, Khanna1, Khanna2}. In this paper, TFD formalism is chosen. It is a real-time finite temperature formalism. In this formalism a thermal state is developed where the main objective is to interpret the statistical average of an arbitrary operator as an expectation value in a thermal vacuum. Two elements are necessary to construct this thermal state: (i) doubling of the original Hilbert space and (ii) use of Bogoliubov transformations. These are two Hilbert spaces, the original space $S$ and the tilde space $\tilde{S}$, are related by a mapping, called the tilde conjugation rules. While the Bogoliubov transformation consists in a rotation involving these two spaces that ultimately introduce the temperature effects.

The Stefan-Boltzmann law and the Casimir effect for the non-abelian GEM field at finite temperature are calculated. The Stefan-Boltzmann law describes the power radiated from a black body in terms of its temperature. The Casimir effect, proposed by H. Casimir \cite{Casimir}, is a quantum phenomenon that appears due to vacuum fluctuations of any quantum field. The results in this case may be at zero or finite temperatures.

This paper is organised as follows. In section II, a brief introduction to the abelian GEM Lagrangian formalism is presented. In section III, an extension to non-abelian GEM field is developed. The energy-momentum tensor associated to non-abelian gauge field is calculated. In section IV, the TFD formalism is introduced. In section V, some applications considering the non-abelian GEM field at finite temperature are analysed. (i) The Stefan-Boltzmann law is calculated. (ii) The Casimir effect at zero temperature is obtained and (iii) the Casimir effect at finite temperature is calculated. In section VI, some concluding remarks are presented.

\section{Lagrangian formulation of Abelian GEM}

In this section an introduction to the lagrangian formulation of abelian-GEM is presented. The GEM field equations, Maxwell-like equations, are
\bea
\partial^i{\cal E}^{ij}&=&-4\pi G\rho^j,\label{01}\\
\partial^i{\cal B}^{ij}&=&0,\label{02}\\
\epsilon^{\langle ikl}\partial^k{\cal B}^{lj\rangle}+\frac{1}{c}\frac{\partial{\cal E}^{ij}}{\partial t}&=&-\frac{4\pi G}{c}J^{ij},\label{03}\\
\epsilon^{\langle ikl}\partial^k{\cal E}^{lj\rangle}+\frac{1}{c}\frac{\partial{\cal B}^{ij}}{\partial t}&=&0,\label{04}
\eea
where $G$ is the gravitational constant, $\epsilon^{ikl}$ is the Levi-Civita symbol, $\rho^j$ is the vector mass density, $J^{ij}$ is the mass current density and $c$ is the speed of light. The quantities ${\cal E}^{ij}$, ${\cal B}^{ij}$ and $J^{ij}$ are the gravitoelectric field, the gravitomagnetic field and the mass current density, respectively. The symbol $\langle\cdots\rangle$ denotes symmetrization of the first and last indices i.e. $i$ and $j$.

The fields ${\cal E}^{ij}$ and ${\cal B}^{ij}$ are expressed in terms of a symmetric rank-2 tensor field, $\tilde{\cal A}$, with components ${\cal A}^{ij}$, such that
\bea
{\cal B}=\textrm{curl}\,\tilde{\cal A}.\label{B}
\eea
and 
\bea
{\cal E}+\frac{1}{c}\frac{\partial\tilde{\cal A}}{\partial t}=-\mathrm{grad}\,\varphi,\label{E}
\eea
where  $\varphi$ is the GEM counterpart of the electromagnetic (EM) scalar potential $\phi$.

Defining ${\cal F}^{\mu\nu\alpha}$ as the gravitoelectromagnetic tensor the GEM field equations become
\bea
\partial_\mu{\cal F}^{\mu\nu\alpha}&=&\frac{4\pi G}{c}{\cal J}^{\nu\alpha},\\
\partial_\mu{\cal G}^{\mu\langle\nu\alpha\rangle}&=&0,
\eea
where ${\cal J}^{\nu\alpha}$ depends on quantities $\rho^i$ and $J^{ij}$ that are the mass and the current density, respectively. In addition, the gravitoelectromagnetic tensor  is defined as
\bea
{\cal F}^{\mu\nu\alpha}=\partial^\mu{\cal A}^{\nu\alpha}-\partial^\nu{\cal A}^{\mu\alpha}
\eea
and the dual GEM tensor is defined as
\bea
{\cal G}^{\mu\nu\alpha}=\frac{1}{2}\epsilon^{\mu\nu\gamma\sigma}\eta^{\alpha\rho}{\cal F}_{\gamma\sigma\rho}.
\eea

Using these definitions, the GEM lagrangian density is given as \cite{Khanna}
\bea
{\cal L}_G=-\frac{1}{16\pi}{\cal F}_{\mu\nu\alpha}{\cal F}^{\mu\nu\alpha}-\frac{G}{c}\,{\cal J}^{\nu\alpha}{\cal A}_{\nu\alpha}.
\eea
This lagrangian allows considering several gravitational applications involving the graviton, such as interactions with other fundamental particles. This makes it possible to study several related topics.

In this way, the GEM theory is described by two fields ${\cal E}^{ij}$ and ${\cal B}^{ij}$, which are symmetric and traceless tensors of second order. These fields can be expressed in terms of the symmetric gravitoelectromagnetic potential $A_{\mu\nu}$ \cite{Khanna, gauge}, analogous to that of electromagnetism $A_{\mu}$. Thus, $A_{\mu\nu}$ is the fundamental field in GEM and naturally it has two indices \cite{Khanna, gauge}.

It is important to note that GEM equations correspond to the weak-field approximation of General Relativity. They do not describe strong fields and therefore, do not include the full Einstein equations. To be more specific, the Abelian GEM corresponds to the linear part of Einstein equations and the non-Abelian GEM corresponds up to the second-order in the weak-field approach.

\section{Non-Abelian GEM}

Let us consider an extension of the GEM field to include the non-abelian gauge transformations \cite{nonGEM}. Then, in this section the lagrangian for the non-abelian GEM field is presented and the energy-momentum tensor associated to the non-abelian field is calculated.

In order to obtain the non-abelian gauge transformation for the GEM field, let us investigate the Dirac lagrangian under global and local gauge transformations. The free Dirac lagrangian is given as
\bea
{\cal L}_D=-i\overline{\psi}(x)\gamma_\mu\partial^\mu\psi(x)+m\overline{\psi}(x)\psi(x),
\eea 
where $\psi(x)$ is a two-component column vector. This lagrangian is invariant under global ${\cal SU}(2)$ gauge transformation given as
\bea
\psi'(x)=U\psi(x)
\eea
with $U$ being a $2\times 2$ unitary matrix that is written as $U=e^{iH}$, where $H$ is a hermitian matrix. To study local gauge transformation, more details are necessary. 

Let us assume that the local gauge transformation is
\bea
\psi'(x)=U(x)\psi(x)=e^{ig H(x)}\psi(x),
\eea
where $g$ is the coupling constant and $H(x)$ is the hermitian $2\times 2$ matrix given by
\bea
H(x)=\mathbf{\sigma}\cdot\mathbf{a}(x)
\eea
with $\mathbf{a}(x)$ being real functions of $x$ and $\mathbf{\sigma}$ are Pauli matrices. The Pauli matrices $\sigma^i$ (i=1,2,3) are the generators of the non-abelian group ${\cal SU}(2)$ satisfying the commutation relations $[\sigma^i, \sigma^j]=2i\epsilon^{ijk}\sigma^k$. In a more compact form,  $\mathbf{a}(x)$ is written as
\bea
\mathbf{a}(x)=p_\alpha \mathbf{b}^\alpha(x)
\eea
where $\mathbf{b^\alpha}(x)$ are vectors associated to each of the four directions in Minkowski space-time and $p_\alpha$ are the components of the one-form $\tilde{p}$. Then the local gauge transformation becomes
\bea
\psi'(x)=e^{igp_\alpha\mathbf{\sigma}\cdot\mathbf{b}^\alpha(x)}\psi(x).
\eea
The Dirac lagrangian is not invariant under this local gauge transformation since the derivative $\partial^\mu\psi'(x)$ introduces a new term in the lagrangian. In order to obtain an invariant lagrangian, a covariant derivative is defined as
\bea
D^\mu=\partial^\mu-igp_\alpha\mathbf{\sigma}\cdot\mathbf{A}^{\mu\alpha}(x),\label{cd}
\eea
where the tensor gauge field $\mathbf{A}^{\mu\alpha}(x)$ has three components $\mathbf{A}^{\mu\alpha}(x)=(A^{\mu\alpha}_1(x), A^{\mu\alpha}_2(x),A^{\mu\alpha}_3(x))$ and it transforms as 
\bea
\mathbf{A'}^{\mu\alpha}_k(x)=A^{\mu\alpha}_k(x)+\partial^\mu b^\alpha_k+2g\epsilon^{ijk}p_\beta A^{\mu\beta}_ib^\alpha_j.
\eea
An important note, there is one tensor gauge field $A^{\mu\alpha}_i(x)$ for each generator $\sigma^i$ of the group ${\cal SU}(2)$. Moreover, in the definition of the covariant derivative $D^{\mu}$ (Eq. (\ref{cd})) the gauge field $A_{\mu\nu}$ should appear to keep the local gauge invariance like in electromagnetism. In order to have it, the one-form $p_{\alpha}$ is introduced \cite{gauge}. The one-form makes the phase function split into phase factors each associated with one of the four directions in space-time. 

Using these results and replacing the derivative $\partial^\mu$ by the covariant derivative $D^\mu$, the Dirac lagrangian is gauge invariant, i.e.,
\bea
{\cal L}_D=-i\overline{\psi}(x)\gamma_\mu D^\mu\psi(x)+m\overline{\psi}(x)\psi(x).
\eea 
In this formulation three new gauge tensor fields are introduced. To write a full lagrangian invariant under local gauge transformation a kinetic term of $\mathbf{A}^{\mu\alpha}(x)$ must be constructed. To do that, an analogue of the electromagnetic tensor $F_{\mu\nu}$ is constructed. For obtaining the antisymmetric third rank tensor of the gauge field let us consider a covariant derivative (\ref{cd}). Then
\bea
[D^\mu, D^\nu]=-igp_\alpha\mathbf{\sigma}\cdot\mathbf{F}^{\mu\nu\alpha},
\eea
where
\bea
F^{\mu\nu\alpha}_k=\partial^\mu A^{\nu\alpha}_k-\partial^\nu A^{\mu\alpha}_k + 2g\epsilon^{ijk}p_\beta A^{\mu\alpha}_i A^{\nu\beta}_j.
\eea
Then the full lagrangian that is invariant under local ${\cal SU}(2)$ gauge transformations is
\bea
{\cal L}_D=-i\overline{\psi}(x)\gamma_\mu D^\mu\psi(x)+m\overline{\psi}(x)\psi(x)-\frac{1}{16\pi}\mathbf{F}_{\mu\nu\alpha}\cdot\mathbf{F}^{\mu\nu\alpha}.
\eea
This lagrangian describes two equal mass Dirac fields interacting with three massless tensor gauge fields.

In conclusion, GEM is an approach based on formulating gravity in analogy to electromagnetism. In this way, GEM becomes a gauge field theory of gravity in contrast with the geometric theory of General Relativity. Then it is expected that the ${\cal SU}(2)$ be the gauge symmetry group. It is the Weyl tensors ${\cal E}^{ij}$ and ${\cal B}^{ij}$ that keep the connection of GEM to gravity.    

Now let us determine the energy-momentum tensor associated with the non-abelian GEM field.

\subsection{Energy-momentum tensor for non-abelian GEM}

Hereafter, the lagrangian density for the free non-abelian GEM field is considered, i.e.,
\bea
{\cal L}=-\frac{1}{16\pi}F^a_{\mu\nu\alpha}F^{\mu\nu\alpha a}
\eea
The  index $a$ is summed over the generators of the gauge group and for an ${\cal SU}(N)$ group one has $ a,b,c=1\cdots N^{2}-1.$ Here, as a first application of the non-abelian GEM, the self-interaction between the tensor gauge fields is ignored.

Using the energy-momentum tensor definition
\bea
T^{\mu\nu}&=&\frac{\partial {\cal L}}{\partial(\partial_\mu A^a_{\lambda\alpha})}\partial^\nu A^a_{\lambda\alpha}-\eta^{\mu\nu}{\cal L},
\eea
the energy-momentum tensor associated with the non-abelian GEM field is
\bea
T^{\mu\nu}=\frac{1}{4\pi}\left[-{\cal F}^{\mu a}_{\lambda\alpha}{\cal F}^{\nu\lambda\alpha a}+\frac{1}{4}\eta^{\mu\nu}{\cal F}^a_{\rho\sigma\theta}{\cal F}^{\rho\sigma\theta a}\right].\label{EMT}
\eea

To avoid a product of field operators at the same space-time point, the energy-momentum tensor is written as
\bea
T^{\mu\nu}(x)&=&\frac{1}{4\pi}\lim_{x'\rightarrow x}\left\{\tau\left[-{\cal F}^{\mu a}_{\lambda\alpha}(x){\cal F}^{\nu\lambda\alpha a}(x')+\frac{1}{4}\eta^{\mu\nu}{\cal F}^a_{\rho\sigma\theta}(x){\cal F}^{\rho\sigma\theta a}(x')\right]\right\}
\eea
where $\tau$ is the time order operator.

The quantization of the non-abelian GEM field requires that
\bea
\pi^{\kappa\lambda a}=\frac{\partial {\cal L}}{\partial(\partial_0 A^a_{\kappa\lambda})}=-\frac{1}{4\pi}{ F}^{0\kappa\lambda a}.
\eea
Adopting the Coulomb gauge, where $A^{0i}=0$ and $\mathrm{div}\tilde{A}=\partial_i A^{ij}=0$, the covariant quantization is carried out and the commutation relation is
\bea
\left[A^{ij a}({\bf x},t),\pi^{kl b}({\bf x}',t)\right]&=&\frac{i}{2}\Bigl[\delta^{ik}\delta^{jl}-\delta^{il}\delta^{jk}-\frac{1}{\nabla^2}\Bigl(\delta^{jl}\partial^i\partial^k-\nonumber\\
&-&\delta^{jk}\partial^i\partial^l-\delta^{il}\partial^j\partial^k+\delta^{ik}\partial^j\partial^l\Bigl)\Bigl]\delta^3({\bf x}-{\bf x}')\delta^{ab}.\label{CR}
\eea
Other commutation relations are zero.

In order to write the energy-momentum tensor, let us consider
\bea
\tau\left[{\cal F}^{\alpha\kappa\gamma a}(x){\cal F}^{\mu\nu\rho a}(x')\right]={\cal F}^{\alpha\kappa\gamma a}(x){\cal F}^{\mu\nu\rho a}(x')\theta(x_0-x_0')+{\cal F}^{\mu\nu\rho a}(x'){\cal F}^{\alpha\kappa\gamma a}(x)\theta(x_0'-x_0),
\eea
with $\theta(x_0-x_0')$ being the step function. In calculations that follow, we use the commutation relation, eq. (\ref{CR}), and 
\bea
\partial^\mu\theta(x_0-x_0')=n^\mu_0\delta(x_0-x_0'),
\eea
where $n^\mu_0=(1,0,0,0)$ is a time-like vector. 

Using these definition the energy-momentum tensor for the non-abelian GEM field becomes
\bea
T^{\mu\nu}(x)=-\frac{1}{4\pi}\lim_{x'\rightarrow x}\left\{\Delta^{\mu\nu,\lambda\epsilon\omega\upsilon}(x,x')\tau\left[A^a_{\lambda\epsilon}(x)A^a_{\omega\upsilon}(x')\right]\right\},
\eea
where
\bea
\Delta^{\mu\nu,\lambda\epsilon\omega\upsilon}(x,x')=\Gamma^\mu\,_{\rho\alpha,}\,^{\nu\rho\alpha,\lambda\epsilon\omega\upsilon}(x,x')-\frac{1}{4}\eta^{\mu\nu}\Gamma^{\rho\sigma\theta,}\,_{\rho\sigma\theta,}\,^{\lambda\epsilon\omega\upsilon}(x,x'),
\eea
with
\bea
\Gamma^{\alpha\kappa\gamma,\mu\nu\rho,\lambda\epsilon\omega\upsilon}(x,x')=\left(g^{\kappa\lambda}g^{\epsilon\gamma}\partial^\alpha-g^{\alpha\lambda}g^{\epsilon\gamma}\partial^\kappa\right)\left(g^{\nu\omega}g^{\rho\upsilon}\partial'^\mu-g^{\mu\omega}g^{\rho\upsilon}\partial'^\nu\right).
\eea

The vacuum expectation value of the energy-momentum tensor  leads to the expression
\bea
\langle T^{\mu\nu}(x)\rangle &=& \langle 0|T^{\mu\nu}(x)|0\rangle,\nonumber\\
&=&-\frac{1}{4\pi}\lim_{x'\rightarrow x}\left\{\Delta^{\mu\nu,\lambda\epsilon\omega\upsilon}(x,x')\left\langle 0\left| \tau\left[A^a_{\lambda\epsilon}(x)A^a_{\omega\upsilon}(x')\right]\right| 0\right\rangle\right\},
\eea
where the graviton propagator is 
\bea
\left\langle 0\left| \tau\left[A^a_{\lambda\epsilon}(x)A^a_{\omega\upsilon}(x')\right]\right| 0\right\rangle&=&\delta^{ab}\left\langle 0\left| \tau\left[A^a_{\lambda\epsilon}(x)A^b_{\omega\upsilon}(x')\right]\right| 0\right\rangle,\nonumber\\
&=&i\delta^{ab}\,D^{ab}_{\lambda\epsilon\omega\upsilon}(x-x'),
\eea
with
\bea
D^{ab}_{\lambda\epsilon\omega\upsilon}=\frac{1}{2}\delta^{ab}\left(g_{\lambda\omega}g_{\epsilon\upsilon}+g_{\lambda\upsilon}g_{\epsilon\omega}-g_{\lambda\epsilon}g_{\omega\upsilon}\right)G_0(x-x'),
\eea
and $G_0(x-x')$ is the massless scalar field propagator. Then the vacuum expectation value of $T^{\mu\nu}(x)$ becomes
\bea
\langle T^{\mu\nu}(x)\rangle=-\frac{3i}{8\pi}\lim_{x'\rightarrow x}\left\{\Gamma^{\mu\nu}(x,x')G_0(x-x')\right\},\label{eqT}
\eea
with
\bea
\Gamma^{\mu\nu}(x,x')=8\left(\partial^\mu\partial'^\nu-\frac{1}{4}\eta^{\mu\nu}\partial^\rho\partial'_\rho\right).\label{Gamma}
\eea

Now the main objective is to study the effects due to temperature and spatial compactification in eq. (\ref{eqT}). To achieve such an objective, the Thermo Field Dynamics formalism is used.

\section{Thermo Field Dynamics (TFD) formalism}

Here the Thermo Field Dynamics (TFD) formalism is introduced. TFD is a quantum field theory at finite temperature \cite{Umezawa1, Umezawa2, Umezawa22, Khanna1, Khanna2}. In this formalism, the statistical average of any operator is equal to its expected value in a thermal vacuum. For this equality to be true, two main elements are required, i.e., (i) doubling of the original Hilbert space and (ii) the Bogoliubov transformation.

This doubling is defined as  ${\cal S}_T={\cal S}\otimes \tilde{\cal S}$, where $\tilde{\cal S}$ and ${\cal S}$ are the tilde and original Hilbert space, respectively. The Bogoliubov transformation  corresponds to a rotation of the tilde and non-tilde variables which introduces the thermal effects. To understand this doubling of Hilbert space, let us consider 
\bea
\left( \begin{array}{cc} d(\alpha)  \\ \tilde d^\dagger(\alpha) \end{array} \right)={\cal B}(\alpha)\left( \begin{array}{cc} d(k)  \\ \tilde d^\dagger(k) \end{array} \right),
\eea
where ${\cal B}(\alpha)$ is the Bogoliubov transformation given as
\bea
{\cal B}(\alpha)=\left( \begin{array}{cc} u(\alpha) & -v(\alpha) \\ 
-v(\alpha) & u(\alpha) \end{array} \right)
\eea
with 
\bea
v^2(\alpha)=(e^{\alpha\omega}-1)^{-1}, \quad\quad u^2(\alpha)=1+v^2(\alpha).\label{phdef}
\eea
The $\alpha$ parameter is the compactification parameter defined by $\alpha=(\alpha_0,\alpha_1,\cdots\alpha_{D-1})$ and $\omega$ is energy. The temperature effect is described by the choice $\alpha_0\equiv\beta$ and $\alpha_1,\cdots\alpha_{D-1}=0$. In this case, with $\alpha=\beta$, the quantities $v^2(\beta)$ and $u^2(\beta)$ are related to the Bose distribution.

In order to introduce an application of TFD formalism, let's consider the free scalar field propagator. Then in a doublet notation it is given as
\bea
G_0^{(ab)}(x-x';\alpha)=i\langle 0,\tilde{0}| \tau[\phi^a(x;\alpha)\phi^b(x';\alpha)]| 0,\tilde{0}\rangle,
\eea
where $\phi(x;\alpha)={\cal B}(\alpha)\phi(x){\cal B}^{-1}(\alpha)$ and $a, b=1,2$. Then
\bea
G_0^{(ab)}(x-x';\alpha)=i\int \frac{d^4k}{(2\pi)^4}e^{-ik(x-x')}G_0^{(ab)}(k;\alpha),
\eea
where 
\bea
G_0^{(11)}(k;\alpha)\equiv G_0(k;\alpha)=G_0(k)+v^2(k;\alpha)[G_0(k)-G^*_0(k)],
\eea
with
\bea
G_0(k)=\frac{1}{k^2-m^2+i\epsilon}
\eea
and
\bea
[G_0(k)-G^*_0(k)]=2\pi i\delta(k^2-m^2).
\eea
The parameter $v^2(k;\alpha)$ is the generalized Bogoliubov transformation \cite{GBT}. It is defined as
\bea
v^2(k;\alpha)=\sum_{s=1}^d\sum_{\lbrace\sigma_s\rbrace}2^{s-1}\sum_{l_{\sigma_1},...,l_{\sigma_s}=1}^\infty(-\eta)^{s+\sum_{r=1}^sl_{\sigma_r}}\,\exp\left[{-\sum_{j=1}^s\alpha_{\sigma_j} l_{\sigma_j} k^{\sigma_j}}\right],\label{BT}
\eea
with $d$ being the number of compactified dimensions, $\eta=1(-1)$ for fermions (bosons), $\lbrace\sigma_s\rbrace$ denotes the set of all combinations with $s$ elements and $k$ is the 4-momentum.

For the doubled notation, the vacuum expectation value of the energy-momentum tensor of the non-abelian GEM is
\bea
\langle T^{\mu\nu(ab)}(x;\alpha)\rangle=-\frac{3i}{8\pi}\lim_{x'\rightarrow x}\left\{\Gamma^{\mu\nu}(x,x')G_0^{(ab)}(x-x';\alpha)\right\}.
\eea

In order to obtain a physical (renormalized) energy-momentum tensor, the standard Casimir prescription is used. Then
\bea
{\cal T}^{\mu\nu (ab)}(x;\alpha)=\langle T^{\mu\nu(ab)}(x;\alpha)\rangle-\langle T^{\mu\nu(ab)}(x)\rangle.
\eea
In this form a measurable physical quantity is given as
\bea
{\cal T}^{\mu\nu (ab)}(x;\alpha)=-\frac{3i}{8\pi}\lim_{x'\rightarrow x}\left\{\Gamma^{\mu\nu}(x,x')\overline{G}_0^{(ab)}(x-x';\alpha)\right\},\label{VEV}
\eea
where 
\bea
\overline{G}_0^{(ab)}(x-x';\alpha)=G_0^{(ab)}(x-x';\alpha)-G_0^{(ab)}(x-x').
\eea

In the next section, some applications for different choices of $\alpha$-parameter are developed.

\section{Some Applications}

In this section applications, that consider the temperature effects and spatial compactifications, are calculated.

\subsection{Stefan-Boltzmann law}

As a first application, consider the thermal effect that appears for $\alpha=(\beta,0,0,0)$. In this case the generalized Bogoliubov transformation becomes
\bea
v^2(\beta)=\sum_{j_0=1}^\infty e^{-\beta k^0 j_0}.
\eea
Then the Green function is given as
\bea
\overline{G}_0^{(11)}(x-x';\alpha)&=&\int \frac{d^4k}{(2\pi)^4}e^{-ik(x-x')}\sum_{j_0=1}^\infty e^{-\beta k^0 j_0}\left[G_0(k)-G_0^*(k)\right],\nonumber\\
&=&2\sum_{j_0=1}^\infty G_0\left(x-x'-i\beta j_0 n_0\right),\label{1GF}
\eea
where $n_0^\mu=(1,0,0,0)$. Then the energy-momentum tensor at finite temperature is
\bea
{\cal T}^{\mu\nu(11)}(x;\beta)&=&-\frac{6i}{\pi}\lim_{x'\rightarrow x}\left\lbrace \sum_{j_0=1}^\infty\left(\partial^\mu\partial'^\nu-\frac{1}{4}g^{\mu\nu}\partial^\rho\partial'_\rho\right) G_0\left(x-x'-i\beta j_0 n_0\right) \right\rbrace.
\eea
Using the Riemann Zeta function, i.e.,
\bea
\zeta(4)=\sum_{j_0=1}^\infty\frac{1}{j_0^4}=\frac{\pi^4}{90},\label{zetaf}
\eea
the Stefan-Boltzmann law for the non-abelian GEM field is obtained as
\bea
E(T)\equiv{\cal T}^{00(11)}(x;\beta)=\frac{\pi}{10} T^4.
\eea
Note that, the energy density of the non-abelian gauge fields is similar to the abelian field case. Here the numeric value is multiplied by the group generators number.

\subsection{Casimir effect at zero temperature}

Here $\alpha=(0,0,0,iL)$ is chosen and the Bogoliubov transformation is
\bea
v^2(L)=\sum_{l_3=1}^\infty e^{-iLk^3l_3}.
\eea
The Green function is
\bea
\overline{G}_0^{(11)}(x-x';L)&=&2\sum_{l_3=1}^\infty G_0\left(x-x'-Ll_3z\right).\label{2GF}
\eea
A sum over $l_3$, for $L=2d$, defines the nontrivial part of the Green function with the Dirichlet boundary condition. With these conditions, the energy-momentum tensor becomes
\bea
{\cal T}^{\mu\nu(11)}(x;d)&=&-\frac{6i}{\pi}\lim_{x\rightarrow x'}\left\lbrace \sum_{l_3=1}^\infty\left(\partial^\mu\partial'^\nu-\frac{1}{4}g^{\mu\nu}\partial^\rho\partial'_\rho\right) G_0\left(x-x'-2dl_3z\right) \right\rbrace.\label{zerotemp}
\eea
For $\mu=\nu=0$ the Casimir energy to the non-abelian field case is
\bea
E(d)&=&{\cal T}^{00(11)}(x;d)=-\frac{\pi}{480d^4}
\eea
and for $\mu=\nu=3$ the Casimir pressure for the non-abelian GEM field is
\bea
P(d)&=&{\cal T}^{33(11)}(x;d)=-\frac{\pi}{160d^4}.
\eea
The negative sign shows that the Casimir force between the plates is attractive, similar to the case of the electromagnetic field and of the abelian GEM field.

\subsection{Casimir effect at finite temperature}

For $\alpha=(\beta, 0, 0,i2d)$ the temperature effects and spatial compactifications are considered. In this case the Bogoliubov transformation becomes
\bea
v^2(k^0,k^3;\beta,d)&=&v^2(k^0;\beta)+v^2(k^3;d)+2v^2(k^0;\beta)v^2(k^3;d),\nonumber\\
&=&\sum_{j_0=1}^\infty e^{-\beta k^0j_0}+\sum_{l_3=1}^\infty e^{-iLk^3l_3}+2\sum_{j_0,l_3=1}^\infty e^{-\beta k^0j_0-iLk^3l_3}.
\eea
The Green function, corresponding to the first two terms, is given in eq. (\ref{1GF}) and in eq. (\ref{2GF}), respectively. The Green function associated with the third term is
\bea
\overline{G}_0^{(11)}(x-x';\beta,d)&=&2\int \frac{d^4k}{(2\pi)^4}e^{-ik(x-x')}\sum_{j_0,l_3=1}^\infty e^{-\beta k^0j_0-iLk^3l_3}\left[G_0(k)-G_0^*(k)\right],\nonumber\\
&=&4\sum_{j_0,l_3=1}^\infty G_0\left(x-x'-i\beta j_0n-2dl_3z\right).\label{3GF}
\eea
Then the Casimir energy and pressure at finite temperatue are given, respectively, by
\bea
E(\beta, d)={\cal T}^{00(11)}(\beta;d)&=&\frac{\pi}{10\beta^4}-\frac{\pi}{480d^4}+\frac{6}{\pi^3}\sum_{j_0,l_3=1}^\infty\frac{3(\beta j_0)^2-(2dl_3)^2}{[(\beta j_0)^2+(2dl_3)^2]^3},\label{ED}\\
P(\beta, d)={\cal T}^{33(11)}(\beta;d)&=&\frac{\pi}{30\beta^4}-\frac{\pi}{160d^4}+\frac{6}{\pi^3}\sum_{j_0,l_3=1}^\infty\frac{(\beta j_0)^2-3(2dl_3)^2}{[(\beta j_0)^2+(2dl_3)^2]^3}.\label{P}
\eea
Note that, the first and second terms are the Stefan-Boltzmann law and Casimir effect at zero temperature, respectively. While the third term correspond to the Casimir effect at finite temperature. In the last case both effects, temperature and spatial compactification, are present.

\section{Conclusion} 

The non-abelian GEM field is investigated. First, the lagrangian formulation for abelian GEM field is presented. Then using the principle of local gauge invariance an extension of the non-abelian GEM field is constructed. The symmetry group for the non-Abelian GEM is the group ${\cal SU}(2)$. The Abelian and non-Abelian GEM have a correspondence with the weak-field approach of General Relativity. The abelian GEM has a structure equivalent to the weak-field approximation of first order and non-Abelian Weyl GEM is equivalent to the weak-field approximation up to second order. For simplicity, the self-interactions terms of the non-abelian gauge field are ignored. Then the energy-momentum tensor is calculated. The TFD formalism is used to introduce thermal effects. This formalism requires two basic ingredients: the doubling of the Hilbert space and the Bogoliubov transformation. With this formalism the vacuum expectation value of the energy-momentum tensor is obtained and thus, some applications for non-abelian GEM field are investigated. The Stefan-Boltzmann law and the Casimir effect at finite temperature are calculated. Our results show that the non-abelian quantities are similar to the abelian quantities. The main difference consists in the fact that the non-abelian results are equal to the abelian result multiplied by the number of gauge fields. These results are similar to the case of the electromagnetic field. For example, the non-abelian GEM Casimir effect is attractive as is the electromagnetic case. In addition, calculations involving ${\cal SU}(2)$ group and GEM have not been done in the literature. This work is the first to introduce this type of approach.

\section*{Acknowledgments}

This work by A. F. S. is supported by CNPq projects 308611/2017-9 and 430194/2018-8.

\end{document}